\documentclass[ag]{copernicus}

\usepackage{color}

\begin{document}

\title{{Electron-cylotron maser radiation from electron holes: \\ Downward current region}}

\author[1,2]{{R. A. Treumann}
}
\author[3]{{W. Baumjohann}}
\author[4]{{R. Pottelette}}

\affil[1]{Department of Geophysics and Environmental Sciences, Munich University, Munich, Germany}
\affil[2]{Department of Physics and Astronomy, Dartmouth College, Hanover NH 03755, USA}
\affil[3]{Space Research Institute, Austrian Academy of Sciences, Graz, Austria}
\affil[4]{LPP-CNRS/INSU, 94107 Saint-Maur des Foss\'es, France}

\runningtitle{Electron-Hole Radiation}

\runningauthor{R. A. Treumann, W. Baumjohann, and R. Pottelette}

\correspondence{R. A.Treumann\\ (rudolf.treumann@geophysik.uni-muenchen.de)}

\received{ }
\revised{ }
\accepted{ }
\published{ }


\firstpage{1}

\maketitle

\begin{abstract}
The electron-cyclotron maser emission from electron holes is applied to holes generated in the downstream current region of the aurora. We suggest that part of the fine structure observed in the auroral kilometric radiation is generated by the electron-cyclotron maser mechanism in the downstream current region. The argument goes that the main background auroral kilometric radiation source is still located in the partial electron ring (horseshoe) distribution of the upward current region while the fine structure is caused by electron holes generated predominantly in the downward current region. Since both regions always exist simultaneously they are acting in tandem in generating auroral kilometric radiation by the same mechanism though in different ways.

\end{abstract}

\introduction
Emission of auroral kilometric radiation has commonly been attributed to the upward current region for reasons of observation of electron distributions with sufficiently steep and positive perpendicular velocity space gradients $\partial f_e(v_\|,v_\perp)/\partial v_\perp>0$ on the electron phase-space distribution $f_e(v_\|,v_\perp)$, viz. loss-cone, ring or horseshoe distributions etc. \citep[more precisely the relativistic momentum distribution, cf., e.g.,][and others; for a review see Treumann, 2006]{wulee1979,winglee1983,pritchett1984a,pritchett1984b,pritchett1984c,louarn2006}. The downward current region has barely been considered a source because none of those distributions is found in that region. The electron component observed here  is known to consist purely of ionospheric electrons which, by a sufficiently strong downward field-aligned electrostatic potential of not yet completely clarified origin, are sucked up from the ionosphere and  become accelerated into an upward  magnetic-field aligned and fairly cold electron beam which carries the downward auroral current \citep{carlson1998}. Such an electron distribution is believed to be inactive with respect to the electron-cyclotron maser; if it causes radiation, then at  the best via a nonlinear wave-wave interaction process like the one known in generating solar type III bursts and emission from the electron foreshock at low harmonics of the electron plasma frequency $\omega_e$. In the auroral region $\omega_e\ll\omega_{ce}$ is sufficiently below the electron cyclotron frequency $\omega_{ce}$ for making this mechanism obsolete.  

On the other hand, downward field-aligned current strengths are large in the auroral downward current region. It has therefore early been understood that the auroral downward current region would be a preferable location for the generation of phase space holes via some current instability, both ion holes \citep{berman1986,tetreault1988,tetreault1991,gray1990,ergun1998a,ergun1998b,ergun1998c} and electron holes \citep{ergun2002}. Ion holes, in that case, are the result of the ion-acoustic instability evolving from ion acoustic solitons in the presence of dissipation or either from lower-hybrid or from electrostatic whistler waves. Their dynamics has been investigated in the above papers for parallel propagation. Electron holes would be excited by the Buneman instability \citep{buneman1958,buneman1959,treumann1997} as in the case of the upstream current region. Their dynamics has been the subject of a large number of papers \citep[][and others]{ergun2002,muschietti1999a,muschietti1999b,newman2001,newman2002,oppenheim2001}. Neither of these structures have been considered to be of any importance in the generation of radiation. Their mentioning in connection to radiation was entirely restricted to the upward current region and to a few papers only \citep{pottelette2000,pottelette2005,treumann2006,treumann2008}. In a recent paper \citep{treumannea2011} we developed the theory of electron-hole maser radiation for the upward current region.

\begin{figure}[t!]
\centerline{{\includegraphics[width=0.5\textwidth,clip=]{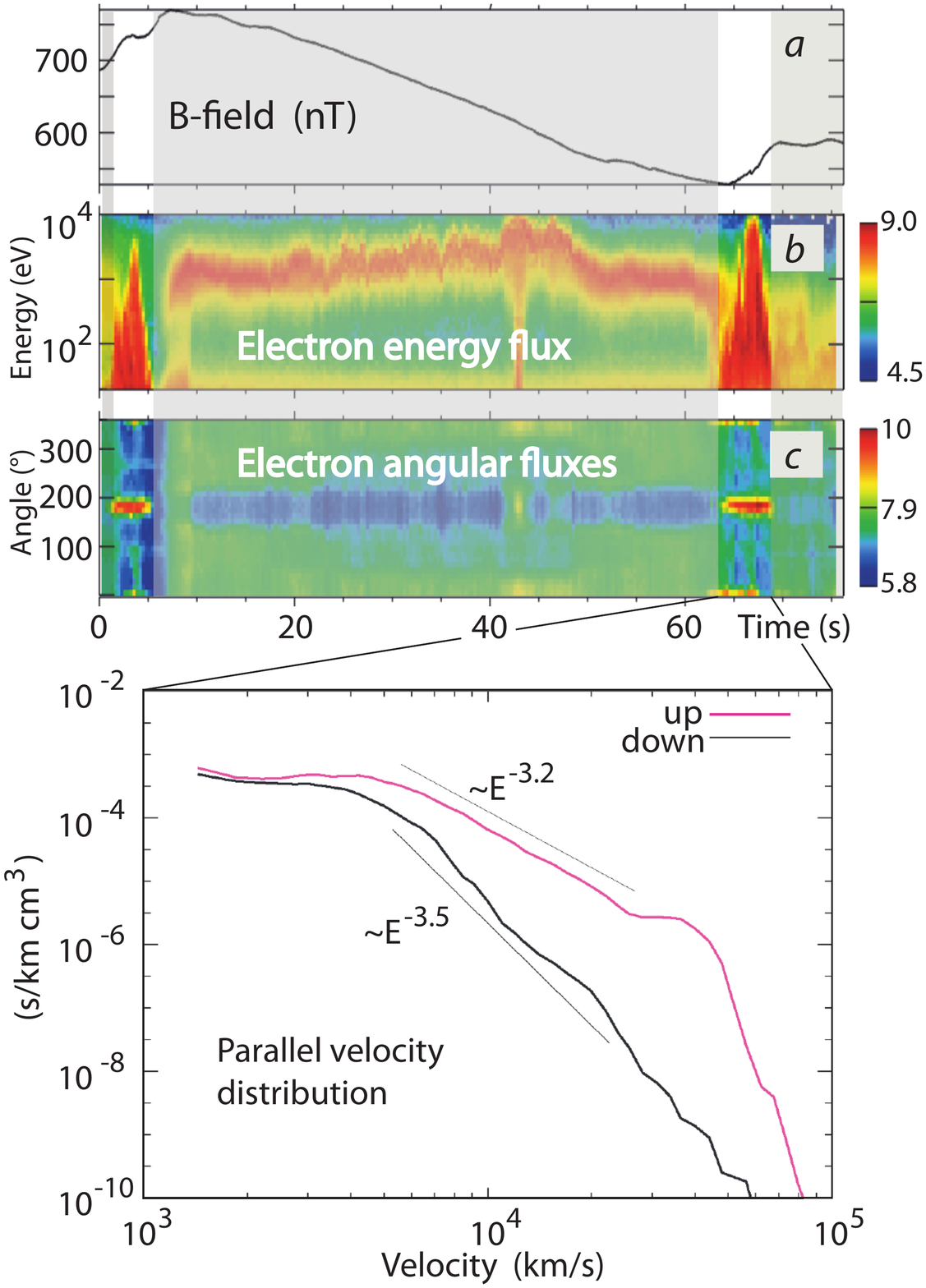}
}}\vspace{-7mm}
\caption[ ]
{\footnotesize {The two downward current regions of the {\small FAST 1773} path. (a) Magnetic field with the two positive gradient regions indicating downward currents. (b) Electron energy flux. The downward current region is characterised by their typical broad electron energy fluxes. (c) Angular distribution of flux. Showing its concentration at angles of 180$^\circ$. The parallel electron distribution is shown for the second region in the lower part of the figure (upward electron velocities are negative, here shown mirrored into positive speeds). The low energy flat part is caused by electric-field acceleration. Upward electron fluxes are substantially higher than downward and exhibit a sharp cut-off at high energies after a power law part, indicating strong upward acceleration. The difference between upward and downward fluxes is responsible for the downward current and corresponds to a non-zero upward bulk velocity (Data obtained within the University of California Berkeley - France cooperation).}}
\vspace{-0.3cm}\label{fig-ray1}
\end{figure}
In the present communication we transform the latter theory to the downward current region. This attempt is stimulated by the discouraging result obtained in the previous paper when applying the electron-hole maser to the upward current region. There we combined high-resolution observations with simulations of electron holes and radiation theory. We found that electron holes can indeed be emitters of narrowband X-mode radiation via the electron-cyclotron maser. The emitted bandwidths and emission frequencies fall all into place when compared with observation. Also, the polarisation of the radiation turned out to be correct, on the X-mode (or Z-mode) branch. However, we were discouraged by the low growth and amplification rates of the radiation. 

In an attempt to deal with this problem we realised that the radiation was sufficiently short wavelength and low frequency for being completely trapped inside the electron hole. When the radiation is generated at the electron hole boundary though outside the hole, then the holes do indeed not play any role in radiation production because the radiation does not stay long enough in contact with the hole for reaching large enough amplitudes. Trapping is, however, suggested when the radiation is generated inside the hole on its inner boundary by the fact that the hole is low density and the radiation cannot escape but is confined for the entire lifetime of the hole. For this time it stays in resonance and is continuously amplified. Nevertheless, the amplification factor is small because of the very low electron density. In addition, after decay of the hole and release of the radiation its frequency is still far below the X-mode cut-off, and the radiation can propagate only in the Z-mode. Though Z-mode radiation has indeed been observed in the auroral kilometric source region, its escape to free space poses another problem \citep[cf., e.g.,][and references therein]{louarn2006}. 

The downward current region has the advantage of hosting much higher electron densities than in the auroral cavity. Still, these densities are low enough for the electron cyclotron maser if only the electron distribution exhibits a sufficiently steep perpendicular velocity gradient. This, however, can be generated locally by electron holes. Any radiation excited by the electron-cyclotron maser inside the holes, being trapped for the lifetime of the hole and afterwards released can then leak out from the downward current region into the auroral cavity, the upward current region where it may contribute to fine structure. Here we show that indeed electron holes in the downward current region are capable of this. Their growth and amplification rates are higher than in the upward current region, and their frequencies are above the upward current region X-mode cut-off.
\begin{figure*}[t!]
\centerline{{\includegraphics[width=0.75\textwidth,clip=]{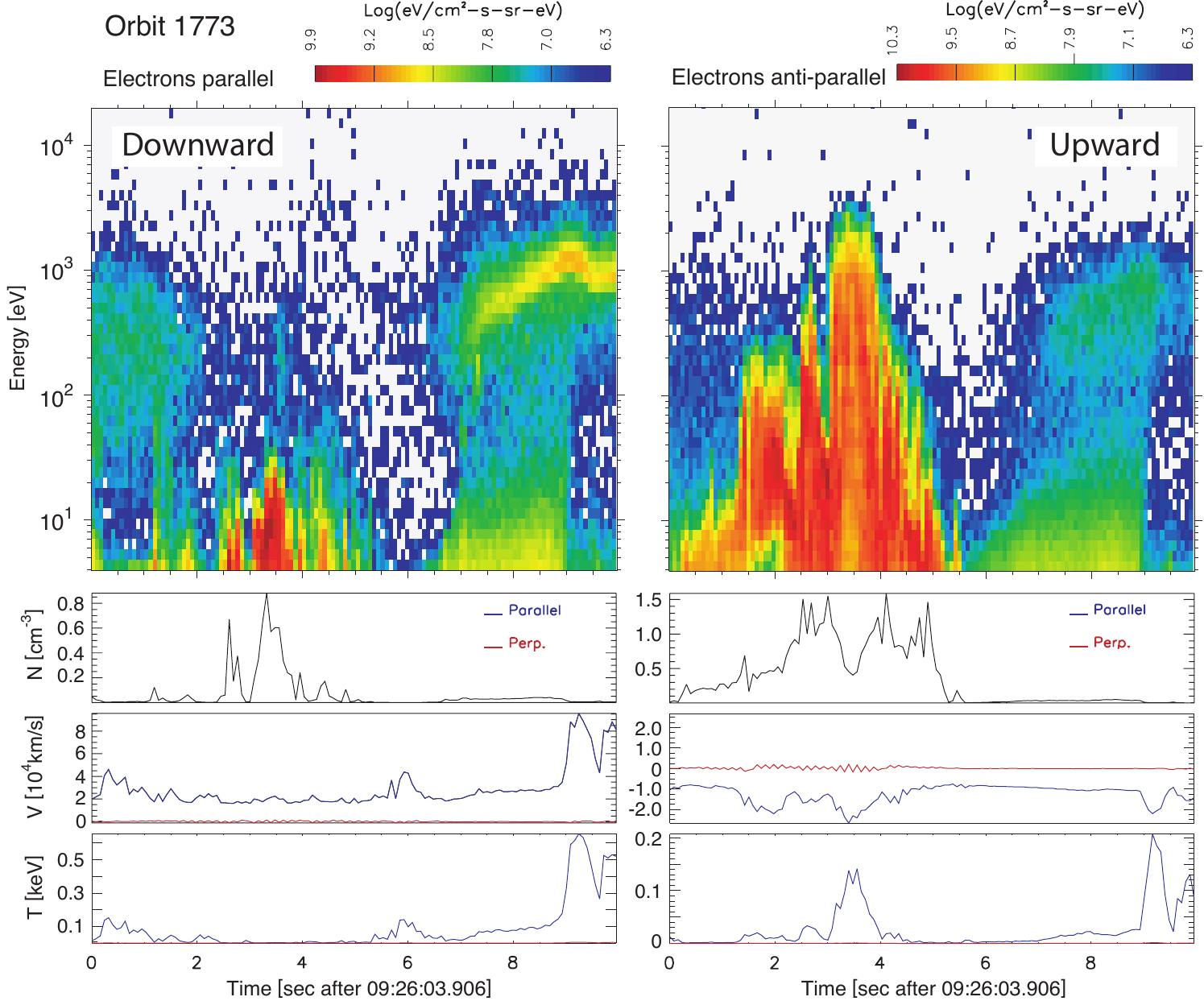}
}}
\caption[ ]
{\footnotesize {A high resolution view of the second downward current event in Fig. \ref{fig-ray1} for downward (left) and upward electron fluxes. Downward fluxes [i.e. the $v_\|>0$ tail of $f_e(v_\|)$] are concentrated at low electron energy around 10 eV with substantial particle density of $\sim$0.6 cm$^{-3}$. Upward fluxes reach densities of $\gtrsim$1 cm$^{-3}$. Thus plasma frequencies in this region range around 10 kHz $\lesssim\omega_e/2\pi \lesssim14$ kHz. Parallel bulk velocities range between $10^4$ km/s $\lesssim V\lesssim1.5\times10^4$ km/s while temperatures maximise at $T\sim150$ eV. Thus the thermal velocities are $v_e<10^4$ km/s. Under these conditions the Buneman instability is excited and electron holes are easily formed.}}\label{fig-edita2}
\vspace{-0.3cm}
\end{figure*}

\section{Hole structure in the downward current region}
Observations in the upward current region have been described exhaustively in \citet{treumannea2011}. We refer to their Figure 3 referring to downward current observations which flank the upward current region data. In the magnetic trace of panel 1 they belong to the positive field gradient for the first 10 s and the 65-80 s of the spacecraft path. Panels 5 and 6 show the electron fluxes in energy and angle, respectively. Electron fluxes maximise at few 10 eV ($\sim$100 eV) and extend up to few keV ($\sim$10 keV). Angular distributions peak strongly at $\sim$180$^\circ$ antiparallel to the local magnetic field direction (angular resolution is $\pm11^\circ$) with weak fluxes at oblique angles. The last two panels show the ions to have energies of few keV and being distributed into conics (very low upward ion fluxes). 

As for an argument in favour of the generation of electron holes we need to show data which support their existence. This is done here just for completeness as evidence for holes and direct measurements of holes have been presented earlier in the above cited literature. Figure \ref{fig-ray1} is an extract of the mentioned Fig. 3 in \citet{treumannea2011}, {\small FAST 1773}. The two narrow unshaded sections show the upward electron flux in the downward current region during this crossing of the auroral region. For the second of these we have produced the time-averaged electron velocity distribution function along the magnetic field separated into upward and downward distributions. These distributions show the excess of upgoing electrons at velocities above $v\sim2\times10^3$ km/s peaking at $v\sim4\times10^3$ km/s before turning into a power law decay for roughly one decade in velocity until  $v\sim3\times10^4$ km/s. This power law decay is indicated as the corresponding power law $\sim{\cal E}^{-3.2}$ in the equivalent parallel energy distribution. It is followed by an exponential cut-off at velocities $v>4\times10^4$ km/s. Just before the cut-off the upward  bump on the distribution indicates the presence of a change in the \emph{average} character of the distribution which resembles a beam bounding the distribution function. Note that the flat top at low speeds is typical for distributions accelerated by stationary electrostatic fields \citep[for the latter see][]{andersson2002}.

We plot in Figure \ref{fig-edita2} the electron data and moments of the distribution for this second downward current region in higher resolution. On the left are the downward electron fluxes, on the right the current-carrying upward fluxes. Only these are of interest. Obviously all electron fluxes exhibit a high temporal dynamic. 

There are several noticeable observations which we briefly list as follows:

--- First, the total electron density of upward and downward electrons amounts to roughly $N\approx 2$ cm$^{-3}$ which corresponds to a plasma frequency of $\omega_e/2\pi\sim 13$ kHz. This value is still far below the local electron cyclotron frequency and emission frequency of auroral kilometric radiation which was in this case $\omega_{ce}/2\pi\sim 483$ kHz \citep{treumannea2011}. Thus the ratio $\omega_e/\omega_{ce}\sim3\times10^{-2}\ll1$ indicates that the electron cyclotron maser radiation theory from electron holes will in principle be applicable in its ordinary version but will lead to much larger hole-maser-growth rates than were obtained in the upward current region. 

--- Second, the upward velocity of the current-carrying electrons reaches values $V>2\times10^4$ km/s with an average value of $\langle V\rangle \sim 1.8\times10^4$ km/s, roughly a factor of $\gtrsim 2$ larger than the electron thermal velocity $v_e<10^4$ km/s. This is sufficiently high for the Buneman instability to be excited. Thus the conditions for generation of electron holes are satisfied in the downstream current region. Ion velocities (not shown) are downward being of the order of few times 100 km/s. This velocity is adding to the downward current velocity thereby improving the above condition. Moreover, the existence of phase space holes, ion holes as well as electron holes, both on the Debye scale, has been confirmed experimentally here \citep{carlson1998,ergun1998a,ergun1998b,ergun1998c}. 

--- Third and most interestingly, inspection of the electron spectrum indicates that the electron fluxes are generally bounded in energy by strongly increased fluxes in a narrow energy band. These are the fluxes which have added up to the beam-like feature on the averaged upstream electron distribution in Figure \ref{fig-ray1}. The instantaneous width in energy -- or temperature -- of this feature can be read from the maximum just before second 4 on the right in Figure \ref{fig-edita2} to amount to $\sim100$ eV (or few times 100 eV) yielding a thermal speed of $v_{eb}\sim 10^3$ km/s for this feature. This feature follows all the variations of the fluxes even down to the narrowest ones seen in this figure. It is highly suggestive to take it as the indication of the cold electron beam which is seen in the simulations \citep{newman2001} of electron holes as the nonlinearly `electron-hole cooled' residuum of the initial warm electron-current `beam' distribution \citep[cf. Figure 7 in][]{treumannea2011}. 

\section{Electron holes in the downstream current region}
The mechanism of generation of electron holes in the downstream current region is the same as that in the upstream current region. The only differences are:

\noindent
--- the absence of a ring-horseshoe electron distribution which is replaced by the upward electron flow distribution in Figure \ref{fig-ray1}, 

\noindent
--- the substantially higher electron plasma frequency, 

\noindent
--- the propagation direction of the electron holes, which by theory is in the same direction as the electron `beam-current' velocity (caused by the Buneman instability the hole speed $V_h<V$ is slightly below the current speed; thus,  in the electron beam frame, the holes move against the beam); in the downward current region it should be upward towards lower electron cyclotron frequencies, and

\noindent
--- the question of the perpendicular extension of the electron holes in velocity space. 

Note again that the extension in configuration space is of secondary importance because, for the electron-cyclotron maser to work, a perpendicular gradient on the electron distribution function in velocity space is absolutely necessary. This latter point is crucial and is of non-trivial nature, as it was in the upward current region. There we provided ample arguments for a perpendicular hole structure in velocity space. These will have to be repeated and modified below.  

Our assumption is that the electron holes will be excited by the Buneman instability \citep{buneman1958,buneman1959,treumann1997}. This requires that, for a given fixed perpendicular velocity the effective electron drift along the magnetic field $v_\|(v_\perp)>v_e(v_\perp)$ should exceed the effective parallel electron thermal velocity at this perpendicular speed in order to excite the Buneman mode. The angular fluxes in Figure \ref{fig-ray1} suggest that there is only a small spread of the distribution into perpendicular direction. However, time, angle and energy resolutions do not permit for any precise conclusion on the scale of the electron holes. From theory and simulation it is known that the extension of the holes along the magnetic field is of the order of $\lesssim 100\,\lambda_D$ with Debye length $\lambda_D\lesssim 50$ m at $T_e\sim 100$ eV and $N\approx 2$ cm$^{-3}$ yielding  a parallel scale of few km. The thermal gyroradius of electrons in this case is $r_\mathit{ce}^\mathit{th}\sim 10$ m (one should note that in Figure \ref{fig-edita2} the perpendicular temperature given in red is unreliable for the measured particle numbers are rather to small). A 1 keV electron (as in Fig. \ref{fig-edita2}) under these conditions has gyroradius $r_\mathit{ce}\sim 30$ m. Hence gyroscale-limited electron holes should be rather elongated along the magnetic field. Simulations in weak fields show instead \citep{newman2002,oppenheim2001}  oblate holes. Observations in the strong auroral magnetic field region  \citep{franz2000} suggest, on the other hand, that they are about spherical in configuration space, though no distinction is made between the upward and downward current regions. This larger spatial extension is possible at the small electron gyroradii if more than one electron gyro-fluxtube is involved into a hole. However, for the electron distribution this should have an affect only if the electrons are heated into perpendicular direction. 

Figure \ref{fig-ray1} suggests that the angular distribution spreads out between 150$^\circ$ and 230$^\circ$, occasionally filling the entire angular range from parallel to perpendicular, though at low flux. From Figure \ref{fig-edita2} we may thus just conclude that the maximum perpendicular electron velocity may vary from downward (parallel) to upward (antiparallel) with energies between $\sim20$ eV and 2 keV, or a velocity range $10^3<v_\perp<1.5\times10^4$ km/s, quite broad enough for allowing a perpendicular hole structure in phase space. We conclude that independent of the shape of the electron hole in configuration space and independent of the strength of the magnetic field -- which determines the gyroradius of the electron and thus the perpendicular spatial extension of the hole --, the perpendicular electron velocity spans a rather large range allowing for the hole in velocity space to extend up to speeds up to which the Buneman instability can be excited at finite $v_\perp$.

\begin{figure}[t!]
\centerline{\includegraphics[width=0.4\textwidth,clip=]{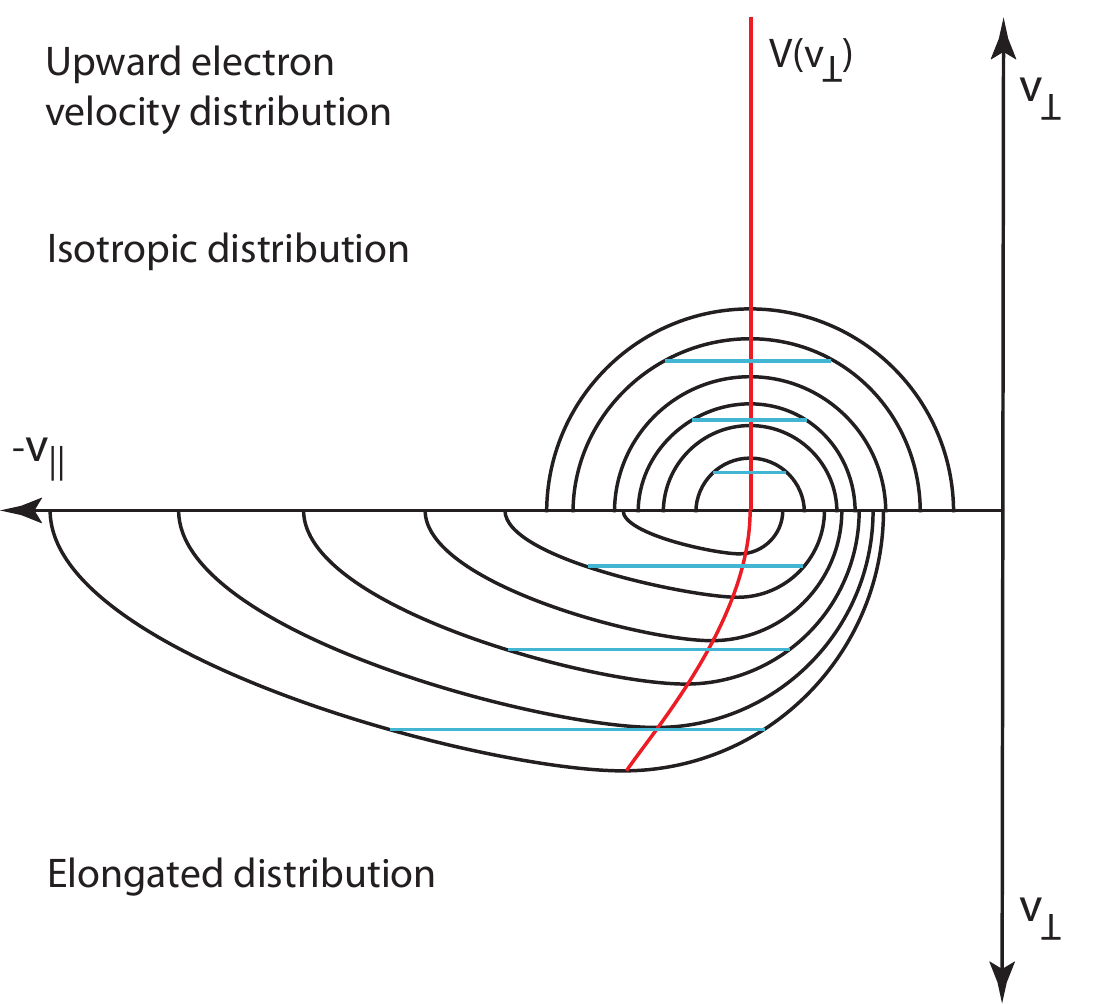}}
\caption[ ]
{\footnotesize Sketch of the upward electron distribution in velocity space. \emph{Upper half plane}: Isotropic shifted distribution. The shift is the (negative) upward electron velocity $V$ (shown in red). The (blue) horizontals indicate the `thermal spread' $v_e(v_\perp)$ at three fixed perpendicular velocities $v_\perp$. \emph{Lower half plane}: Same for an elongated electron distribution (as implied by the measured parallel electron distribution in Figure \ref{fig-ray1}). In this case both current speed $V(v_\perp)$ and `thermal spread' $v_e(v_\perp)$ increase at a larger amount with increasing $v_\perp$ than in the isotropic case.}
\label{fig-sketch}
\vspace{-0.3cm}
\end{figure}

To find out up to which angles (or perpendicular speeds) this will happen, we assume -- for simplicity -- an  isotropic temperature yielding a Maxwellian distribution that is shifted in upward velocity at the amount of the downward current speed $V$ and has parallel temperature $v_e$. This might not be entirely correct as Fig. \ref{fig-ray1} suggests that the parallel electron distribution is elongated, which will introduce some correction to the symmetric assumption but is secondary for the main purpose of this investigation.  
Since $V$ is constant the `effective parallel thermal velocity' is very easy to determine. The result (Fig. \ref{fig-sketch}) is that the perpendicular velocity for which the Buneman condition is satisfied is limited by
\begin{equation}
v_\perp<V\Big(1-\frac{v_e^2}{V^2}\Big)^\frac{1}{2} 
\end{equation} 
or an angle
\begin{equation}
\theta<\mathrm{sin}^{-1}\Big(1-\frac{v_e^2}{V^2}\Big)^\frac{1}{2}
\end{equation}
These conditions are simpler than in the case of the ring-horseshoe distribution of the upward current region \citep{treumannea2011} because the current velocity $V$ does not change with angle to the magnetic field. 
However, due to the elongation of the distribution along the magnetic field this condition will be slightly modified. It will in fact become more stringent because the `effective parallel thermal speed' increases with increasing $v_\perp$ as is obvious from the sketch in Figure \ref{fig-sketch}. This, however, is of little importance for our general straightforward conclusion that a limited range of perpendicular velocities exists in velocity space for which the distribution at given $v_\perp$ will become Buneman unstable and electron holes can be generated. We may thus conclude that within this limiting angle the electron hole extends into perpendicular velocities in phase space. Since the hole is characterised by a depletion of electrons inside the hole and a snowplough effect on the surrounding distribution in the course of which a fast, dense and cold electron beam is created, the hole self-consistently produces a perpendicular velocity gradient in the distribution function at its phase space boundary.

\section{Radiation}

With these explicit preliminaries in mind we can straightforwardly apply the electron-cyclotron maser theory of electron holes developed in our previous communication \citep{treumannea2011}. 

\subsection*{\small{Growth rate}}
Instead of the ring distribution we have in this case the isotropic dense upstreaming electron distribution which at electron velocities outside the hole is bounded by accelerated electron beams of density $N_{eb}$. Working in the hole frame and applying to the circular hole case the resonance condition becomes a circle of radius
\begin{equation}
R_\mathit{res}=\sqrt{2(1-\nu_\mathit{ce})}
\end{equation}
with $\nu_\mathit{ce}=\omega/\omega_{ce}$ and $\omega_{ce}=eB/m_e$ the non-relativistic electron cyclotron frequency. The wave propagates in the X-mode strictly perpendicular to the magnetic field in this simplified case of a circular hole. The maximum growth rate is obtained at maximum resonance
\begin{equation}
R_\mathit{res}^m\approx U-\frac{(\Delta u)^2}{U}
\end{equation}
where $U=V/c, \Delta u=v_e/c$. This yields a maximum growth rate 
\begin{equation}
\mathrm{Im}(\nu_\mathit{ce})\approx \left(\frac{\alpha\pi^3}{8}\right)\left(\frac{\omega_e^2}{\omega_{ce}^2}\right) \bigg(\frac{\omega_{ce}}{\omega}\bigg)
\end{equation}
The plasma frequency is based on the total electron density $N$, while $\alpha=N_{eb}/N\sim 10^{-2}$ is the ratio of the fast accelerated electron beam to background densities. The precise number of $\alpha$ is not known; the above number has been chosen referring to Figure \ref{fig-ray1} assuming that the enhancement in the distribution near cut-off of the distribution is caused by the presence of those beams. It is thus an average number which might be somewhat larger for single holes up to, say, one order of magnitude at most. Choosing the conservative number of $\alpha\sim 1\%$ only we believe that we are on the safe side. With measured frequencies $\omega_e/2\pi\approx 15$ kHz, $\omega_{ce}/2\pi\approx 480$ kHz we have
\begin{equation}
\mathrm{Im}(\omega)\,\gtrsim\, 2\times10^{-4} \bigg(\frac{\omega_{ce}}{\omega}\bigg)\frac{\omega_{ce}}{2\pi}
\end{equation}
The emission frequency $\omega\lesssim\omega_{ce}$ is very close to the electron cyclotron frequency such that it does not contribute substantially to the growth rate. Thus, due to the higher background density this value is roughly two orders of magnitude higher than the corresponding value of the hole-maser growth rate obtained in the upstream current region. Electron-cyclotron maser emission from electron holes in the downstream current region is thus much stronger than from the upstream current region. Still, the linear growth rate is small and thus the radiation weak unless it becomes further amplified; since, however, in the downstream current region no other source of the electron cyclotron maser exists, any electron cyclotron-maser radiation originating in this region is probably exclusively due to the presence of electron holes in the downstream current region. 

\subsection*{\small{Amplification}}
Electron-cyclotron maser radiation from electron holes is weak not only in the upstream but also in the downstream current region unless it undergoes further amplification. Only then can it explain any of the observed fine structures in the auroral kilometric radiation \citep[or maser radiation from other planets, the sun and other astrophysical objects, cf., e.g.][and references therein]{zarka2005,hess2009,mottez2010,treumann2011a,treumann2011}. Radiation that would be generated outside the electron hole in the perpendicular velocity gradient between the hole and the fast accelerated beam or between the beam and the background distribution will not experience amplification. Radiation of this kind escapes from its source and becomes trapped only in the large-scale auroral cavity where it is reflected between the walls of the cavity until escaping to free space. During its propagation it gets readily out of resonance with the radiation source barely encountering anymore any favourable perpendicular velocity gradient for being further amplified. It thus may contribute to a weak radiation background only. 

In order to experience substantial amplification, the electron-cyclotron maser radiation produced by phase space holes must necessarily be trapped inside the hole for staying long enough in resonance in order to achieve large amplification rates. Such a radiation is generated inside the hole and, because of its frequency being below the X-mode cut-off in the hole, is trapped for the entire life-time of the hole. By the maser theory, its frequency is below the local electron cyclotron frequency. It also cannot escape to the hole environment because of the even higher external X-mode cut-off frequency. It bounces back and forth between the hole boundaries. Trapping of the radiation we have also proposed for radiation from electron holes in the upstream current region \citep{treumannea2011}. 

Electron hole life times have been suggested to be limited by the transverse instability of the hole \citep{oppenheim2001,newman2002,ergun2002} while depletion of the macroscopic distributions is mostly due to very low frequency waves \citep{labelle2002}. Simulations fixed the life times of holes against such transverse instability to $1500< t\omega_e\lesssim3000$ \citep{newman2002}. This corresponds to $t\omega_{ce}\gtrsim 3.6\times10^5$ for the {\small{FAST 1773}} observations. Compared to the growth time obtained above this implies roughly $\sim70$ e-foldings of the wave during complete trapping, or a huge -- and thus unrealistically high --  amplification factor in our case. We may thus realise that radiation trapping inside the hole might be a viable process for keeping the radiation in resonance and amplifying it over the lifetime of the hole until being released from the hole. But trapping cannot be complete because the amplification would become excessively large. 

In the downstream current region lifetimes of the above length correspond to times 0.5 s$<t<$1 s and to propagation distances of $5\times 10^3$ km $<L_h< 10^4$ km of the hole along the magnetic field at nominal hole velocity $V_h\lesssim V\sim 10^4$ km/s. These distances are also somewhat large in the auroral region in addition to yielding unreasonably high amplification factors of the order of $10^{13}$ or so. Roughly ten times shorter lifetimes would cause lower amplification rates but are in disagreement with simulations. Moreover, for trapping of wavelengths of the order of 1 km the perpendicular size of the holes should be of same order. This is suggested by the observations of about symmetrical holes with comparable parallel and perpendicular dimensions \citep{franz2000} but is theoretically not confirmed yet.  As mentioned before, the extension is up to few kilometers along the magnetic field. The perpendicular extension is not precisely known. It may range from 100 m to few km, depending on the symmetry of the hole. Shorter perpendicular sizes cause severe problems because the wave frequency is fixed to $\omega<\omega_{ce}$, not permitting for large perpendicular wavenumbers $k_\perp\sim \omega/c$ and shorter than $\sim$km wavelengths. Waves at these frequencies are inhibited to propagate outside the hole where they become evanescent. Though it would be possible for the instability that the hole pumped energy into the wave only over the fraction of wavelength inside the hole and that most of this energy is absorbed on the fraction outside the hole. However, this is a speculation the confirmation of which requires an eikonal solution of the long wave amplification problem. 

On the other hand, accepting long lifetimes and assuming symmetric holes that (in configuration space), it may be concluded that any excessive \emph{amplification must be reduced} by some mechanism like \emph{self-absorption} of the radiation inside the hole during trapping time. This is indeed \emph{possible for holes in the downstream} current region. These holes move up along the magnetic field from regions of strong magnetic fields into regions of low magnetic fields. We have shown \citep{treumann2011a} that the interior of electron holes is capable of absorption of wave energy at a weaker rate than emission. This is due to the hole being filled with a low density hot electron component the phase space density of which exhibits negative perpendicular velocity gradients $\partial f_\mathit{trapped}/\partial v_\perp<0$. The frequency of absorption is higher than the frequency of emission, $\omega_\mathit{abs}>\omega$. Motion of the hole during its lifetime from higher to lower magnetic fields shifts the emission, amplification and absorption slowly to ever lower frequencies. By this a wave which had been emitted at some frequency $\omega$ in the high $\omega_{ce}$ region shifts out of resonance and amplification entering into the absorption band where it becomes damped, though at a slower rate. A mechanism of this kind might limit the amplification of the radiation and reduce its intensity to observed values.  

One may get an impression of the downward resonant frequency shift assuming a geomagnetic dipole field $B=B_0/R^3$ with $B_0$ the magnetic field at Earth's surface $r=R_\mathrm{E}$, and $R=r/R_\mathrm{E}=(1+\Delta s)$ -- neglecting any latitudinal variation over the short distance $\Delta s=V_h\Delta t$ (measured in  $R_\mathrm{E}$) a hole travels at speed $V_h$ during lifetime $\Delta t$. The resonance radius $R_\mathit{res}=$ const is fixed for the given electron distribution. Thus, the resonance frequency varies as the magnetic field, i.e. $\Delta \omega\approx -3\omega_{ce}(V_h/R_\mathrm{E})\Delta t$. At speed $V_h$ few times $10^3$ km/s this implies for the {\small FAST 1773} passage a substantial decrease in resonant emission frequency of $\Delta\omega\sim 5\%\sim 20$ kHz. A wave amplified during the first phase of trapping is quickly shifted in frequency out of resonance into the absorption range thus reducing the amplification by the required order of magnitude from $\sim70$ e-foldings to $\lesssim10$ e-foldings, i.e. to a reasonable  amplification factor of $\lesssim10^4$ in wave amplitude and $\sim10^8$ in wave power. 

A quantitative estimate of the amplification/absorption rate is hardly possible without gathering more knowledge about the real velocity of the hole, its extension, the trapped number and energy densities, and the location of the holes. Previously \citep[][based on their Fig. 2]{treumannea2011} we determined the background noise (whether instrumental or natural remained unclear) of the radiation power to ${\cal P}\sim 10^{-12}$ W. The maximum radiated power in the fine structures was found from observation to be of order ${\cal P}\sim 8.5\times10^{-8}$ W, corresponding to an amplification of $\sim10^4$ only. These numbers were obtained in the upward current region. Adopting the philosophy of the present paper that the fine structure would be generated in the downward current region one needs to explain the lower amplification rate. This may be due to at least three factors:
 
\noindent
--- a much lower natural downward current region radiation-background noise level than in the upward current region; this seems reasonable because in the downward current region no mechanism is known which would be capable of generating of electromagnetic background noise at the auroral kilometric radiation frequencies. Any background must have leaked in from the upward current region as residual radiation produced from the ring-horseshoe distribution. Its intensity will thus be diluted;

\noindent
--- a higher absorption rate inside the hole; this is not unreasonable as the density of the hole-trapped electron component might be larger than assumed, its temperature might be larger than assumed, the shift of the radiation out of resonant amplification might be larger than assumed;

\noindent
--- the attenuation of radiation with distance from the boundary of the downward current region to the observation site in the upward current region; this is also reasonable because the intensity of the strictly perpendicular radiation should decay with distance as $r^{-2}$. In this case the radiation intensity in fine structure should increase when approaching the downward current region from the upward current region side, which is not obvious from the data \citep[Fig. 3 of][]{treumannea2011}.  Maybe the distances are to short for this effect to be important.

\subsection*{\small{Radiation escape}}
In the above scenario electron holes of appropriate size in the downward current region can indeed generate, amplify and self-absorb narrow band radiation when moving up along the magnetic field from high to lower magnetic field strengths at velocity of few thousand kilometres per second for a limited lifetime. 

Acknowledging this fact, one is confronted with the additional problem of how the radiation escapes from the holes and why it is observed mostly in the upstream and not in the downstream current region. This observational distinction is, though, difficult to make because both regions are embedded into the auroral cavity and localisation of the microscopic radiation source is problematic if not impossible. 

Hole-trapped and amplified radiation can escape from the hole once the hole decays and has moved up to magnetic field strengths low enough for the trapped radiation to have frequency above the X-mode cut-off. The condition for this to happen is easy to check for the given ratio $\omega_e/\omega_\mathit{ce}\approx1/20=0.05$ in the upstream region, yielding an X-mode cut-off frequency 
\citep[e.g.,][p. 228\,ff]{treumann1996}  
\begin{equation} 
\omega_\mathrm{X-co}\approx\omega_\mathit{ce}(1+\omega_e^2/\omega_\mathit{ce}^2) \approx 1.003\ \omega_\mathit{ce}
\end{equation}
which is only slightly higher than the upper-hybrid frequency. At $\omega_\mathit{ce}/2\pi=480$ kHz the cut-off is just $\sim 1$ kHz above the electron cyclotron frequency. It is therefore no problem for any radiation that is trapped in the hole to find itself in free space above the local X-mode cut-off after the upward moving hole has moved up sufficiently high on the field line. In fact, this kind of untrapping might limit the amplification rate even more than any of the above noted restrictions. In this case untrapping would  determine the limitation and ultimate intensity of the electron-hole maser radiation in the downstream current region.

The radiation, being emitted perpendicular in the downstream current region, easily leaks out from it into the even lower density upstream current region. There it will be observed as the upstream current region is substantially more extended than the former. On the other hand, any holes near the boundaries of the downstream current region will anyway release most of their trapped and amplified radiation directly to the upstream current region because the downstream current region is narrow and  bounded by the magnetic field, and X-mode radiation from the about circular holes propagates perpendicular to the magnetic field.

\conclusions
With these remarks on the escape of radiation from the hole we conclude this brief communication on the relevance of electron hole-maser emission in the downstream current region. 

We stress that the downstream current region turns out to be as important as the upstream current region to the emission of auroral kilometric radiation by the electron-cyclotron maser mechanism. While the upstream current region, because of the presence of loss cone and ring-horseshoe distributions in the electron component may dominate the generation of the auroral kilometric background continuum radiation, production of fine structure is probably added from the downstream current region by the processes of the kind described -- or similar ones involving other microscopic structures like double layers, solitons or possibly even ion holes, all producing density gradients. The condition is that they also produce positive perpendicular velocity space gradients on the local electron distribution. 

The downstream current region does not provide electron velocity distributions with large-scale positive velocity space gradients. Such gradients can, however, be caused on microscopic scales by electron holes, solitons or double layers. We have shown in the previous paper \citep{treumannea2011} and in this brief communication that such gradients will be the rule whenever electron holes (or similar structures) evolve. We have provided arguments for electron hole generation by the Buneman instability and have demonstrated that such gradients turn the electron holes into emitters and absorbers of electron-maser radiation. In the upstream current region our model faced some hardly solvable problems. However, in the downstream current region the problems became less severe. We have offered some reasonable solutions. Growth rates for radiation obtained their were high compared to the upstream current region. 

Problems arose with amplification and trapping of radiation inside holes. Trapping is possible only for holes of comparable parallel and perpendicular sizes. In this case very high and thus unreasonable amplification rates were obtained. However, this problem turned out to be less restrictive as self-absorption of radiation during upward motion of the holes helps keeping the amplification low. This we have discussed only qualitatively. The process itself awaits a quantitative treatment. If it can be solved positively then release of radiation after amplification from holes into the upstream current region will be responsible for the observed fine structure in auroral kilometric radiation. 

We note finally that such a model might also be applicable to other planets, the sun, magnetised stars, and a some astrophysical objects like non-relativistic and relativistic shocks \citep{treumann2009,treumann2011} as well where they would indicate the presence of strong magnetic field-aligned electric fields and potentials.

\begin{acknowledgements}
This research was part of an occasional Visiting Scientist Programme in 2006/2007 at ISSI, Bern. RT thankfully recognises the assistance of the ISSI librarians, Andrea Fischer and Irmela Schweizer. He appreciates the encouragement of Andr\'e Balogh, Director at ISSI. The FAST spacecraft observations used have been obtained within the University of California at Berkeley -- France cooperation. The data in Fig. \ref{fig-ray1} have previously been published. They are reprinted here (with changes) with the permission of the American Geophysical Union. Fig. \ref{fig-edita2} is based on an analysis of those data. Some of the data work was done more than ten years ago with the help and technical efforts of Edita Georgescu which we gratefully acknowledge at this place. RT also acknowledges the sceptical insisting comments of the anonymous referees of the former paper \citep{treumannea2011} on electron holes in the upstream current region which led to the re-thinking of a possible contribution of the downstream current region to auroral kilometric maser radiation.
\end{acknowledgements}

\end{document}